\documentclass[aps,prb,twocolumn,superscriptaddress]{revtex4-2}
\usepackage{amsmath}
\usepackage{amssymb}
\usepackage{epsfig}
\usepackage{color}
\usepackage[colorlinks]{hyperref}
\usepackage{graphicx,amsmath,amssymb,bbold}
\usepackage{graphicx,amsmath}

\hypersetup{colorlinks,citecolor=blue,linkcolor=blue,urlcolor=blue}

\begin{document}

\title{Carrier-induced ferromagnetism in 2D magnetically-doped semiconductor
structures}
\author{V. A. Stephanovich}
\email{stef@uni.opole.pl}
\author{E. V. Kirichenko}
\author{G. Engel}
\affiliation{Institute of Physics, Opole University, Opole, 45-052, Poland}
\author{Yu. G. Semenov}
\email{ygsemeno@ncsu.edu}
\affiliation{Department of Electrical and Computer Engineering, North Carolina State University, Raleigh, North Carolina 27695, USA}
\affiliation {V. Lashkaryov Institute of Semiconductor Physics of National Academy of Sciences of Ukraine, 41 Nauky prospekt, Kyiv 03680, Ukraine}
\author{K. W. Kim}
\email{kwk@ncsu.edu}
\affiliation{Department of Electrical and Computer Engineering, North Carolina State University, Raleigh, North Carolina 27695, USA}
\affiliation{Department of Physics, North Carolina State University, Raleigh, North Carolina 27695, USA }

\begin{abstract}
We show theoretically that the magnetic ions, randomly distributed in a
two-dimensional (2D) semiconductor system, can generate a ferromagnetic
long-range order via the RKKY interaction. The main physical reason is the
discrete (rather than continuous) symmetry of the 2D Ising model of the spin-spin
interaction mediated by the spin-orbit coupling of 2D free carriers, which
precludes the validity of the Mermin-Wagner theorem. Further, the analysis
clearly illustrates the crucial role of the molecular field fluctuations as
opposed to the mean field. The developed theoretical model describes the desired
magnetization and phase-transition temperature $T_c$ in terms of a single
parameter; namely, the chemical potential $\mu$. Our results highlight a path way
to reach the highest possible $T_c$ in a given material as well as an
opportunity to control the magnetic properties externally (e.g., via a gate
bias). Numerical estimations show that magnetic impurities such as Mn$^{2+}$
with spins $S=5/2$ can realize ferromagnetism with $T_c$ close to room
temperature.
\end{abstract}

\maketitle

\affiliation{Institute of Physics, Opole University, Opole, 45-052, Poland}

\affiliation{Department of Electrical and Computer Engineering, North
Carolina State University, Raleigh, North Carolina 27695, USA}%
\affiliation{V. Lashkaryov Institute of Semiconductor Physics of National
Academy of Sciences of Ukraine, Kyiv 03680, Ukraine }

\affiliation{Department of Electrical and Computer Engineering, North
Carolina State University, Raleigh, North Carolina 27695, USA}
\affiliation{Department of Physics, North Carolina State University,
Raleigh, North Carolina 27695, USA }


\section{Introduction}

In the nascent era of spintronics, the studies of localized impurity spins
in the low-dimensional systems have become increasingly important \cite
{fzds2004,rashba2002,bibes}.
At the large inter-spin distances (as compared to a lattice constant), the coupling
between magnetic impurities in metals and semiconductors is primarily due to
the indirect Ruderman-Kittel-Kasuya-Yosida (RKKY) exchange interaction
via free electrons and holes (see Ref.\ \cite{har79} and references therein).
The indirect character of this interaction manifests itself in the fact that
the actual coupling occurs via Friedel oscillations of the free-carrier
charge density in a host material (see, e.g., Refs. \cite{ziman, har79}).
Accordingly, it is very sensitive to the details of the electronic energy
spectrum and spatial dimensionality of the problem. The manipulation
of electronic spectrum parameters such as the energy gap, spin-splitting at the
different points of the Brillouin zone, and the spin-orbit interaction constant can generate
nonstandard collective properties in the impurity spin ensemble (e.g., the
long-range ferromagnetic (FM) ordering), leading potentially to a range of
optoelectronic, spintronic, and energy harvesting applications
\cite{fzds2004,kirilyuk10,tmd1,tmd2}. For instance, the indirect exchange
interaction, mediated by near-surface electrons, was shown to couple local
spins and facilitate the spatial spin correlations \cite{spp}.

Naturally, the spin density generated by an impurity magnetic moment in a
two-dimensional (2D) electron or hole gas can act on another impurity moment
or their cluster such that the resulting collective state becomes very
complex. This complexity can be well captured phenomenologically in terms of
the Landau-Lifshitz-Gilbert (LLG) equation \cite{ll1935,gilb}. The LLG
equation can describe the systems with different long-range magnetic orders
(i.e., FM, antiferromagnetic, helical, etc.) and corresponds to the mean-field
approximation (MFA). From the microscopic point of view, the MFA
amounts to the average of the internal magnetic field over the different
impurity spin configurations (with respect to their indirect exchange
interaction), which is identical for each magnetic ion (i.e., no spatial
fluctuations). This mean field generates the spatially uniform charge carrier
and magnetic ion magnetizations. Subsequent splitting in their mutual spin
spectra is sustained at temperatures $T<T_c$, where $T_c$ is the
FM phase-transition temperature \cite{pary}.

While the MFA is valid for sufficiently large magnetic ion concentrations
$n_{i}$ (e.g., $n_{i}k_{F}^{3}\sim 1$ in 3D systems, where $k_{F}$ is the
Fermi wavevector \cite{dietl1, dietl,pary}), the composition and spin
fluctuations in the magnetic ion ensemble can become substantial at
smaller densities (more precisely, smaller $ n_{i}k_{F}^{3}$ for 3D),
leading eventually to its failure. This physical picture
indicates qualitatively that at a given $n_{i}$, there should exist a critical
free charge-carrier concentration $n_e$ (an areal density in the 2D case)
such that at $n_{e}<n_{e,cr}$, the phase-transition temperature $T_{c}$
becomes zero and the long-range FM order ceases to exist. As the 2D Fermi
wavevector $k_{F}=(2\pi n_{e})^{1/2}$ is related to the free carrier
density, $n_{e,cr}$ can be well expressed through $k_{F}$ and then through
the Fermi energy $E_{F}$ in a parabolic energy band with an effective mass $%
m^{\ast }$. Moreover, the constant density of states in the 2D case leads to the essential
equivalence of $E_{F}$ and the chemical potential $\mu $ when the underlying electron gas
is sufficiently degenerate. This conveniently permits us to use $\mu $ as a control parameter for the
manipulation of FM order characteristics (like local magnetization, spin
polarization of charge carriers, etc.) in the 2D semiconductor structures.
Unlike the metallic counterparts, $n_{e}$ (and thus $\mu $) in a dilute
magnetic semiconductor (DMS) \cite{dietl1, kossut, cibscalb} can be
controlled independently of $n_{i}$ via a number of methods (such as an
external bias or additional doping), highlighting its versatility in
applications.

The purpose of the present paper is to analyze theoretically the effect of
the random distribution of magnetic impurities on the formation of the
long-range FM order in the 2D DMS structures. The geometric
confinement of the structures under consideration enables the application of
the Ising model for the spin-spin interaction of the magnetic impurities
when it is mediated by the free carriers experiencing a spin-orbital field
directed normal to the 2D plane. Our analysis based on the RKKY formalism
clearly illustrates that randomizing the spin-spin interaction results in
the gradual suppression (down to complete elimination) of magnetic order at
the relatively short periods of Friedel oscillations compared to the mean
inter-ion distance (thus, in the regime of high carrier concentrations).
Similarly, it is also revealed that the thermal distribution of the free
carriers makes the FM order impossible at/below the low values of the
chemical potential. These findings clearly indicate the existence of a
limited range of free carrier densities favorable for the FM order unlike in
the MFA. The investigation further highlights the optimum conditions to
achieve the maximum critical temperature $T_c$. A numerical calculation is
provided by using a DMS quantum well (QW) as an example along with a brief
discussion on another magnetically doped 2D system, namely, the few-layered
van der Waals materials.

\section{Theoretical model}

As discussed above, it is convenient to express everything in terms of the
chemical potential $\mu $. Since $\mu $ is directly proportional to $n_{e}$
($\mu \approx E_{F}\sim n_{e}$), the problem can be classified into
two regimes. The first corresponds to a small charge carrier concentration,
where the spatial dependence of the 2D RKKY potential (see below) is
unimportant. Thus, the mean-field treatment can be used. Moreover, the MFA in this case is
well described by the Kondo-like Hamiltonian averaged over the spin states
of localized spin moments \cite{SRMFA}. As $n_{e}$ grows, the Friedel
oscillations of the free carrier density become important, causing the
fluctuations in the magnetic ion subsystem and subsequently precluding the
application of the simple (essentially single-impurity) Kondo-like approach.
The collective behavior of the magnetic ions can instead be described by the
random Ising Hamiltonian with the exchange energy $J(r)$ \cite%
{semst02,semst03} in the form of the 2D RKKY interaction.

We begin with the case of a relatively small $n_{e}$, corresponding to the
transition from a nondegenerate 2D carrier gas to a degenerate one. Here,
the Kondo-like Hamiltonian of the carrier-ion exchange interaction takes the
usual form
\begin{equation}
\mathcal{H}_{K}=\frac{I}{N_{0}}\sum_{j}\mathbf{S}_{j} \cdot \mathbf{s}~ \delta (%
\mathbf{r}-\mathbf{R}_{j}) , \label{1}
\end{equation}
where $I$ is the carrier-ion exchange constant (in units of energy, characterizing the
confinement effect in our 2D structure), $N_0$ is the areal density of cation sites, and
$\mathbf{S}_{j}$ denotes the impurity spin at site $j$ (positioned at $\mathbf{R}_{j}$ in a
host lattice) interacting with an itinerant spin $\mathbf{s}$ at location
$\mathbf{r}$. To be specific, let us apply $\mathcal{H}_{K}$
to the lowest heavy-hole subband, which is separated from the light-hole band
due to the spin-orbit interaction. As the latter interaction quantizes the spin along
the direction normal to the 2D plane (say, the $z$ axis), a fictitious spin operator
$S^*=\pm 3/2$ represents the carrier spin in the basis of heavy-hole eigenfunctions
\cite{R2002}. This transformation leads to the effective Hamiltonian
\begin{equation}
\mathcal{H}_{em}=\frac{I}{3N_{0}}\sum_{j}S_{j,z}S_{z}^{\ast }\delta (\mathbf{%
r}-\mathbf{R}_{j}) , \label{2}
\end{equation}%
where the interaction is reduced to the coupling of spin $z$-components, i.e., the Ising form
of the carrier-ion exchange interaction.  Note that the interaction with light holes
may modify Eq.\ \eqref{2} involving the terms proportional to the
transversal spin components. However, their contributions can be neglected
when the separation between two hole subbands is sufficiently larger than the
thermal energy.  Further,  $\mathcal{H} _{em}$ can be made to resemble the Kondo Hamiltonian
in Eq.\ \eqref{1} by defining the valence band spin operator $S_{e}$ as $ \frac{1}{3}S_{z}^*$.

The \textit{mean-field treatment} of the Hamiltonian $\mathcal{H}_{em}$ starts from
the introduction of mean free carrier $\langle S_{e}\rangle $ and magnetic
ion $\langle S_{i}\rangle $ spin polarizations. Supposing a simple heavy-hole
band structure with an isotropic in-plane effective mass $m^{\ast }$, the
mean carrier-spin polarization $\langle S_{e}\rangle $ can be written in
terms of the spin subband hole populations $n_{\pm }$ with a chemical
potential  $\mu $ as
\begin{equation}
\langle S_{e}\rangle =\frac{1}{2}\frac{n_{+}-n_{-}}{n_{+}+n_{-}},
\label{mik4}
\end{equation}%
where
\begin{equation}
n_{\pm }=\sum_{\mathbf{k}}\frac{1}{e^{(\varepsilon _{\mathbf{k}_{\pm }}-\mu
)/T}+1}.  \label{mik5a1}
\end{equation}
By convention, temperature $T$ is expressed in units of energy. A finite spin
polarization $\langle S_{e}\rangle $ arises due to a finite polarization $%
\langle S_{i}\rangle $ of localized spins, which induces a Zeeman-like
energy of the homogeneous Weiss field modifying the energy of free carriers
with 2D wavevector $\mathbf{k}$
\begin{equation}
\varepsilon _{\mathbf{k}_{\pm }}=\frac{\hbar ^{2}k^{2}}{2m^{\ast }}\pm \frac{%
1}{2}Ix_{i}\langle S_i\rangle.  \label{mik5b}
\end{equation}
Here, $x_{i}=n_{i}/N_{0}$ (i.e., the fraction of impurity magnetic ions in the
host lattice) and $\langle S_i\rangle $ is the thermally averaged  impurity spin
$S_{j,z}$.

Equations \eqref{mik4}-\eqref{mik5b} describe the
dependence of $\langle S_{e}\rangle $ on $\langle S_{i}\rangle $. To determine
the phase-transition temperature $T_c$, at which the infinitesimal magnetization appears,
the expression for $\langle S_{e}\rangle $ needs to be linearized in $\langle S_{i}\rangle$.
This yields
\begin{equation}
\langle S_{e}\rangle =\frac{Ix_{i}}{2T\left( 1+e^{-\xi }\right) \ln \left(
1+e^{\xi }\right) }\langle S_{i}\rangle ,  \label{mik6}
\end{equation}%
where $\xi =\mu /T$. Similarly, the linear approximation for
$\langle S_{i}\rangle$ results in
\begin{equation}
\langle S_{i} \rangle
=\frac{S(S+1)}{3\pi }\frac{m^{\ast }I}{\hbar ^{2}N_{0}}%
\ln (e^{\xi }+1)\langle S_{e}\rangle ,  \label{mik7}
\end{equation}%
where $S$ denotes the spin state of the magnetic impurities.
The above set of relations [i.e., Eqs.\ \eqref{mik6} and \eqref{mik7}] describe the mutual
influence of carrier and magnetic-ion spin polarizations which become
nonzero below a certain critical temperature $T_{c}$. The condition for $T_{c}$ can also
be obtained from Eqs.\ \eqref{mik6} and \eqref{mik7} as
\begin{equation}
T_{c}\left( 1+e^{-\mu /T_{c}}\right) =T_{0},\ T_{0}=\frac{S(S+1)}{6\pi }%
\frac{m^{\star }}{\hbar ^{2}N_{0}}I^{2}x_{i}.  \label{mic8}
\end{equation}%
As shown, $T_{0}$ is a characteristic temperature (in energy units) which depends on the type of a
host material and magnetic impurities. It actually corresponds to the FM
phase-transition temperature in the limit of high carrier densities $\mu \gg T_{0}$ in the
present mean-field treatment
(thus, with no consideration of the fluctuations in the magnetic impurity ensemble). The transcendental
equation for $T_{c}$ given in Eq.\ \eqref{mic8}  can be solved as a function of $\mu $ and $%
T_{0}$ numerically.

It is instructive to compare Eq.\ \eqref{mic8} with $ T_{c}^{3d}$ obtained for a 3D DMS with the
corresponding  volume density of cation sites  $N_{0}^{3d}$ \cite{SRMFA}:
\begin{equation}
T_{c}^{3d}=\frac{S(S+1)}{3}\left( \frac{I}{N_{0}^{3d}}\right) ^2\frac{\chi
_{e}^{(1)}}{g^{2}\mu _{B}^{2}}n_{i}n_{e},  \label{4}
\end{equation}%
where $\chi _{e}^{(1)}$ denotes the carrier magnetic susceptibility per
particle and $g$ and $\mu _{B}$ stand for the Land\'{e} $g$-factor and Bohr
magneton, respectively. Applying  $\chi _{e}^{(1)}=(3/8)g^{2}\mu
_{B}^{2}/E_{F}$ for the degenerate carriers allows us to estimate the ratio
\begin{equation}
\frac{T_{0}}{T_{c}^{3d}}=2\left( \frac{\pi }{3}\right)
^{1/3}(x_{e}^{3d})^{-1/3}  \label{5}
\end{equation}%
in terms of the 3D carrier density $x_{e}^{3d}=n_{e}^{3d}/N_{0}^{3d}$
provided that the other parameters in 3D and 2D cases coincide. 
Since the free carrier
density $n_{e}^{3d}$ is normally much smaller than $N_{0}^{3d}$ in a realistic DMS
sample (e.g., $x_{e}^{3d} < 10^{-3}$) \cite{dietl1}, $T_{c}^{3d}$ is likely to be significantly
lower than $T_0$.  With $T_{c}$ approaching $T_0$ as discussed above [Eq.\ \eqref{mic8}], the confinement
in a 2D DMS system appears to provide a clear advantage over the 3D counterpart.

Now let us turn to the second case of degenerate charge carrier gases. To
account for explicitly the disorder in the magnetic impurity subsystem,
it is convenient to eliminate the charge-carrier spin variables from Eq.\ \eqref{2} in favor of
an effective spin-spin interaction between the localized spin moments.
By following the well-known procedure \cite{har79, ziman, kittel, SRMFA}, this can be achieved with
Eq.\ \eqref{2} rewritten in terms of an effective Ising-like Hamiltonian
\begin{equation}
\mathcal{H}=\sum_{j<j^{\prime }}J(\mathbf{r}_{ij})S_{zj}S_{zj'},
\label{mu0}
\end{equation}%
where $J(\mathbf{r}_{ij})\equiv J_{ij}$ is the interaction potential in
energy units. The Hamiltonian in Eq.\ \eqref{mu0}
contains two "sources of randomness". The first is the thermal disorder,
which means that the spin has a random projection on the specific $i$-th site of a
2D host lattice. Likewise, the spatial disorder is the second source as the
spin can be randomly present or absent at a host lattice site. It's worth
noting that our formalism works for any form of $J_{ij}$ so that the effects
like spin splitting at the corners of the Brillouin zone in some 2D crystal
monolayers (see, e.g., Ref.\ \cite{tmd1} and references therein) can easily
be incorporated.

The indirect interaction of localized spins in a metallic host is usually thought
of in the RKKY form \cite{har79,kittel}. In the bulk semiconductors
with degenerate electron/hole gases, such an interaction results in the FM
ordering \cite{dietl1,dietl,semst02,semst03}. While the particularities of
the electronic band structure in a specific 2D substance can certainly
influence the form of $J_{ij}$ (see, e.g., Ref.\ \cite{abol}), these details
are neglected in demonstrating the universal features as it does not change
the results qualitatively.
In the simplest case of a one-band carrier structure, the RKKY interaction
in 2D can be expressed as \cite{litvdug,bem}
\begin{equation}
J(r)=-U_{0}\bigg[J_{0}(x)Y_{0}(x)+J_{1}(x)Y_{1}(x)\bigg],\ x=k_{F}r\,,
\label{fas1}
\end{equation}%
where $J_{0,1}$ and $Y_{0,1}$ are Bessel and Neumann functions of the zeroth
and first order, respectively \cite{abr}, and
\begin{equation}
U_{0}=J_{0}x_{e},\ J_{0}=\frac{m^{\ast }I^{2}}{4\pi \hbar ^{2}N_{0}},\ x_{e}=%
\frac{n_{e}}{N_{0}}.  \label{fas2}
\end{equation}%
$I$ and $N_{0}$ are as defined earlier Eq.\ \eqref{1}.

\begin{figure}[tbp]
\begin{center}
\includegraphics [width=0.45\textwidth]{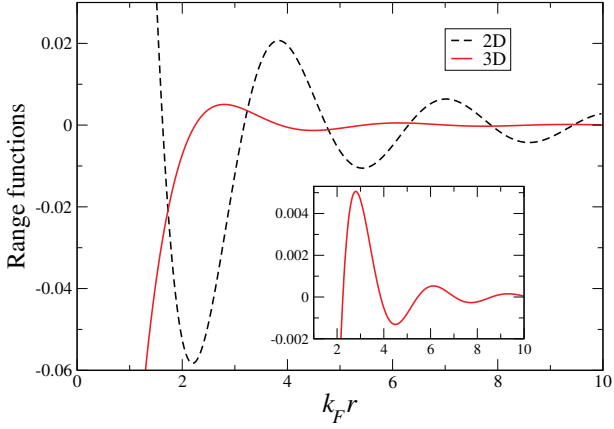}
\end{center}
\par
\caption{Comparison of the range functions for the RKKY interaction
potentials in the 2D (black, dashed line) and 3D (red, solid line) spatial
dimensions. The inset provides a magnified view of the 3D RKKY range
function.}
\label{fig:fi1}
\end{figure}

As the impurity ferromagnetism has already been studied for bulk
semiconductors \cite{semst02,semst03}, it is illustrative to compare the
properties of 2D and 3D range functions that essentially determine the
macroscopic characteristics in the present treatment (such as the
FM phase-transition temperature). The 3D RKKY potential reads
\begin{equation}
J_{3D}(r)=J_{03D} (x_{e}^{3d})^{4/3}F(2k_{F}r),\ F(x)=\frac{x\cos x-\sin x}{x^{4}} ,
\label{fas3}
\end{equation}%
where $J_{0,3D}=I^{2}m^{\ast } (N_{0}^{3d})^{-2/3}(3/\pi )^{1/3}(3/2\hbar ^{2})$.
This expression clearly has a
form much simpler than that in Eq.\ \eqref{fas1}. At a small $x$, the range
function $F(x)\approx -1/3x$, i.e., it is divergent. At a large $x$, on
the other hand, the range function decays like $x^{-3}\cos x$,  which is
rather rapid. In comparison, the asymptotics of the 2D range function at the
small and large values of $x$ become \cite{abr}
\begin{eqnarray}
\frac{J(r)}{U_{0}} &\approx &-\frac{2}{\pi }\ln \frac{x}{2},\ x<<1
\label{fa1} \\
\frac{J(r)}{U_{0}} &\approx &\frac{\sin 2x}{\pi x^{2}},\ x>>1.  \label{fa2}
\end{eqnarray}%
It can be shown that at the lower end of $x$, the 2D range function has a
weaker, logarithmic divergence than that in the 3D case ($\sim 1/x$).
Similarly, the decay to zero at the other end is also slower in the case of
the 2D range function. Figure \ref{fig:fi1} provides a numerical evaluation
of these functions for the full range of $x=k_Fr$. As expected,
the 3D range function decreases much faster than its 2D counterpart at a
large $x$. More specifically, its amplitude at $x>4$ is approximately ten
times smaller than that for the 2D range function. The observed weaker
divergence ($x\rightarrow 0$) and slower decay ($x\rightarrow \infty $)
(thus, the enhanced indirect exchange interaction) is the condition
desirable for a higher FM phase-transition temperature, indicating further the
potential advantage of the 2D structures over the 3D systems. This fact also follows
from quantitative estimation of Eq.\ \eqref{5}.

With an explicit form of the interaction in place [Eq.\ \eqref{fas1}], we
are now in a position to take advantage of the random-field method that has
initially been developed for bulk 3D samples \cite{semst02, semst03}. In
this approach, any spin $S_{zj}$ is treated as a source of random field $%
H_{zi}\equiv \sum_{j\neq i}J(\mathbf{r}_{ij})S_{zj}$ which acts on other
similar spins. Then, all observable properties of the system are determined
by the distribution function $f(H)$ of the random field $H$. More precisely,
any spin average $\overline{\langle A \rangle}$ has the form $\int A(H)f(H)dH$, where the
bar stands for the averaging over the spatial disorder. In addition, $A(H)$ is
a single-particle thermal average with an effective form of the
Hamiltonian ${\cal H}$ \cite{semst02, semst03},
\begin{equation}
{\cal H}_{\mathrm{eff}}=\sum_{i}H_{zi}S_{zi} . \label{eem}
\end{equation}%
The explicit expression for the distribution function $f(H)$ reads
\begin{equation}
f(H)=\left\langle \overline{\delta \left( H-\sum_{j(\neq i)}J(\mathbf{r}%
_{ij})S_{zj}\right) }\right\rangle .  \label{fHf}
\end{equation}%
As the configurational averaging (i.e., over the spatial disorder) and the thermal
averaging cannot be achieved exactly in Eq.\ \eqref{fHf}, we apply an alternative
approach, i.e., the
self-consistent averaging in the spirit of the statistical theory of
magnetic resonance line shape \cite{ston}. By using the spectral
representation of the $\delta $ function, a set of self-consistent equations
can be obtained for the spin averages
$m_{l}=(-1)^{l}~\overline{\langle S_{z}^{l} \rangle}$, $l=1,2,$... (analogous to the $l$-th order 
moment in a sense).
The macroscopic magnetization $\mathcal{M}$ simply becomes $\mathcal{M}=g\mu _{B}m$ 
(with $m \equiv m_1$). 

For an arbitrary spin $S$, the explicit form of this set reads \cite%
{semst02,semst03}
\begin{subequations}
\begin{eqnarray}
&&f(H)=\frac{1}{2\pi }\int_{-\infty }^{\infty }e^{iH\rho +\mathcal{G}(\rho
)}d\rho ,  \label{epq7} \\
&&\mathcal{G}(\rho )=\left\langle \overline{n_{i}\int \left( e^{-iJ(\mathbf{r%
})\sigma \rho }-1\right) d^{2}r}\right\rangle   \nonumber \\
&&=\sum_{\sigma =1/2}^{S}\bigg\{a_{\sigma }\mathcal{F}_{0}(\sigma \rho
)+ib_{\sigma }\mathcal{F}_{1}(\sigma \rho )\bigg\},  \label{epq7a} \\
&&a_{\sigma }+ib_{\sigma }=\int_{-\infty }^{\infty }\bigg[A_{\sigma
}(h)+iB_{\sigma }(h)\bigg]f(H)dH,  \label{epq7b} \\
&&\mathcal{F}_{0}(z)+i\mathcal{F}_{1}(z)=n_{i}\int [\exp \left( iJ(\mathbf{r}%
)z\right) -1]d^{2}r,  \label{epq7c} \\
&&A_{\sigma }(h)+iB_{\sigma }(h)=\frac{2}{Z_{S}}[\cosh (\sigma h)+i\sinh
(\sigma h)],  \label{epq7d} \\
&&Z_{S}=\sum_{\sigma =-S}^{S}e^{-\sigma h}=\frac{\sinh [(S+1/2)h]}{\sinh
(h/2)};h=\frac{H}{T}.  \label{epq7e}
\end{eqnarray}%
The self-consistency is achieved by inserting Eq.\ \eqref{epq7} into Eq.\ %
\eqref{epq7b} and integrating over $H$. This yields
\end{subequations}
\begin{subequations}
\begin{eqnarray}
&&a_{\sigma }+ib_{\sigma }=\int_{-\infty }^{\infty }\left[ \mathcal{A}%
_{\sigma }(z)+i\mathcal{B}_{\sigma }(z)\right]   \notag \\
&&\times \exp \bigg[\sum_{\sigma =1/2}^{S}\bigg\{a_{\sigma }\mathcal{F}%
_{0}(z_{\sigma })+ib_{\sigma }\mathcal{F}_{1}(z_{\sigma })\bigg\}\bigg]dz,
\label{opia} \\
&&\mathcal{A}_{\sigma }(z)+i\mathcal{B}_{\sigma }(z)=\frac{1}{2\pi }%
\int_{-\infty }^{\infty }[A_{\sigma }(h)+iB_{\sigma }(h)]  \notag \\
&&\times \exp (izh)dh,\ z_{\sigma }=\frac{\sigma z}{T}.  \label{eq14b}
\end{eqnarray}%
The above equations are valid for Ising spin of arbitrary magnitude $S$.
Below we apply these equations to the representative case of spin 1/2 as
well as S=5/2. A typical example is Mn ions which are ubiquitous magnetic
impurities in 2D and 3D DMSs (see Refs.\ \cite{tmd1,tmd2,dietl1} and
references therein).

For $S=1/2$, we have $\sigma =\pm 1/2$ and the governing equations result in
the dimensionless magnetization
\end{subequations}
\begin{equation}
m=\int_{-\infty }^{\infty }\tanh \left( \frac{H}{2T}\right) f(H)dH,
\label{s12a}
\end{equation}%
where $f(H)$ is defined by Eq.\ \eqref{epq7} with
\begin{subequations}
\begin{eqnarray}
\mathcal{G}(\rho ) &=&\mathcal{F}_{0}\left( \frac{\rho }{2}\right) +i%
\mathcal{F}_{1}\left( \frac{\rho }{2}\right) ,  \label{s12b} \\
\mathcal{F}_{0}\left( \frac{\rho }{2}\right)  &=&2\pi n_{i}\int_{0}^{\infty }%
\bigg[\cos \left( J(r)\frac{\rho }{2}\right) -1\bigg]rdr,  \label{s12c} \\
\mathcal{F}_{1}\left( \frac{\rho }{2}\right)  &=&2\pi n_{i}\int_{0}^{\infty
}\sin \left( J(r)\frac{\rho }{2}\right) \ rdr.  \label{s12d}
\end{eqnarray}%
Then, Eq.\ \eqref{s12a} assumes the form
\end{subequations}
\begin{equation}
m=T\int_{0}^{\infty }\frac{e^{\mathcal{F}_{0}(\rho )}}{\sinh \frac{\pi \rho T%
}{2}}\sin \left[ m\mathcal{F}_{1}(\rho )\right] d\rho   \label{nam12}
\end{equation}%
following the integration over $H$ \cite{semst03}. This expression defines
the dimensionless magnetization $m$ in a self-consistent manner.

The MFA asymptotics of Eq.\ \eqref{nam12} corresponds to $n_{i}\rightarrow
\infty $ \cite{semst03}, which reduces to $\rho \rightarrow 0$ as well as $%
\mathcal{F}_{0}(\rho )\rightarrow 0$. In this case, we can obtain from Eq.\ %
\eqref{nam12}
\begin{eqnarray}
m &=&T\int_{0}^{\infty }\frac{\sin \left[ m\rho W_{0}\right] }{\sinh \frac{%
\pi \rho T}{2}}d\rho ,  \label{nam13a} \\
W_{0} &=&2\pi n_{i}\int_{0}^{\infty }rJ(r)dr\equiv T_{0}.  \label{nam13b}
\end{eqnarray}%
Here, $W_{0}$ actually corresponds to $T_{0}$ defined earlier
in Eq.\ \eqref{mic8}, which is the Curie temperature $T_{c}$ in the
degenerate regime based on the so-called homogeneous Weiss field
approximation \cite{har79}.  
This coincidence between the results of two
different approaches  is not accidental.
It actually stems from the fact that the RKKY model implicitly takes into account the first-order
contribution in the carrier-ion exchange coupling (i.e., the homogeneous Weiss field) along with
the fluctuating second-order component to ensure the convergence of integral
over the carrier wavevectors \cite{SRMFA}.  As such, the
average over the RKKY oscillations [see the integration in Eq.\ \eqref{nam13b}] cancels out
the second-order term, leaving the contribution from the first-order intact.

Evaluation of the integral in Eq.\ \eqref{nam13a} (i.e., the MFA asymptotics) yields, as expected,
the well-known expression of the mean-field magnetization for the spin 1/2 Ising model
\begin{equation}
m=\tanh \frac{mT_{0}}{T}.  \label{nam14}
\end{equation}%
To obtain the phase-transition condition from Eq.\ \eqref{nam14},
we apply the usual procedure $m\rightarrow 0$, which generates once more $
T=T_{0}$. This procedure can be regarded as a consistency check for our
approximation.

The same procedure $m \to 0$, when applied to the more accurate relation of Eq.\
\eqref{nam12}, leads to the following random-field expression for $T_c$
\begin{equation}  \label{me4}
1=T_c\int_0^\infty \frac{\mathcal{F}_1(\rho )e^{\mathcal{F}_0(\rho )}\ d\rho
}{\sinh \frac{\pi \rho T_c}{2}}.
\end{equation}
Contrary to the MFA shown in Eq.\  \eqref{nam14}, this relation indicates the
existence of a critical condition associated with the case $T_c=0$. More
specifically, at $T_c \to 0$, Eq.\ \eqref{me4} can be reduced to
\begin{equation}
\frac 2\pi \int_0^\infty \frac{\mathcal{F}_1(\rho)}{\rho}e^{\mathcal{F}%
_0(\rho )}\ d\rho =1.  \label{me4a}
\end{equation}
The resulting condition is a complex function of $n_i$ and $n_e$ (thus, $\mu$).
For a given host material and the magnetic ion density $n_i$, it specifies the
free carrier concentration beyond which the long-range order is impossible in the system
even at zero temperature.

The situation for $S=5/2$ is qualitatively the same, while the derivations
are much more cumbersome. In fact, it is difficult even to write a closed
form expression for the dimensionless magnetization. After some algebra, we
arrive at the following equation for $T_{c}$
\begin{subequations}
\begin{eqnarray}
1 &=&\frac{2T_{c}}{7}\int_{0}^{\infty }\mathcal{F}_{11}(\rho )e^{\mathcal{F}%
_{01}(\rho )}L_{5/2}\left( \pi \rho T_{c}\right) d\rho ,  \label{tcu1} \\
\mathcal{F}_{01}(\rho ) &=&\frac{2\pi n_{i}}{3}\int_{0}^{\infty }\bigg(\cos
\frac{5}{2}\zeta +\cos \frac{3}{2}\zeta +\cos \frac{1}{2}\zeta -3\bigg)rdr,
\notag \\
\mathcal{F}_{11}(\rho ) &=&2\pi n_{i}\int_{0}^{\infty }\bigg(\sin \frac{5}{2}%
\zeta +\frac{3}{5}\sin \frac{3}{2}\zeta +\frac{1}{5}\sin \frac{1}{2}\zeta %
\bigg)rdr,  \notag \\
\zeta  &=&\rho J(r).  \label{tcu2}
\end{eqnarray}%
Here $L_{5/2}(z)=\coth \frac{z}{6}-\coth z$.  The expression for the critical concentration
can be derived from Eq. \eqref{tcu1} via the asymptotic relation $L_{5/2}(z\rightarrow
0)=5/z$ to yield
\end{subequations}
\begin{equation}
\frac{10}{7\pi }\int_{0}^{\infty }\frac{\mathcal{F}_{11}(\rho )}{\rho }e^{%
\mathcal{F}_{01}(\rho )}d\rho =1.  \label{tcu3}
\end{equation}%
Equations \eqref{tcu1} and \eqref{tcu3} are solved numerically in Sec.\ III.

\section{Results and Discussion}

For the numerical calculation of the phase-transition temperature, it is
convenient to express both $T_{c}$ and $\mu $ in units of $T_{0}$. In these
units, Eq.\ \eqref{mic8} assumes the form
\begin{equation}
y\left( 1+e^{-\frac{\xi _{0}}{y}}\right) =1,  \label{lwx1}
\end{equation}%
where $y=T_{c}/T_{0}$ and $\xi _{0}=\mu /T_{0}$.
Equation \eqref{lwx1} indicates $y\to 1$ as $\xi_0\gg 1$; i.e., $T_{c}$
cannot exceed $T_{0}$. Furthermore, $T_{c}$ appears to attain its asymptotic value by
$\mu \approx 5T_{0}$.
Being based on the mean-field treatment, the dependence $y(\xi _{0})$
in Eq.\ \eqref{lwx1} is expected to remain valid up to a
moderately degenerate carrier gas in a 2D semiconducting host (obviously
including the nondegenerate case). By contrast, Eq.\ \eqref{me4} or %
\eqref{tcu1} can be used to describe $y(\xi _{0})$ for high values of $\xi
_{0}$ (thus, $\mu $). Assuming magnetic ions Mn$^{2+}$ with $S=5/2$ as
an example, the numerical solution of Eq.\ \eqref{tcu1} similarly shows that
$T_{c} \approx T_0$ at $\mu \approx 5T_{0}$ when approaching from the
opposite, heavily degenerate regime. Thus, the chemical potential around
$5T_{0}$ comprises the crossover between the two $n_{e}$ regimes, enabling
the interpolation between them.

\begin{figure}[tbp]
\begin{center}
\includegraphics [width=0.45\textwidth]{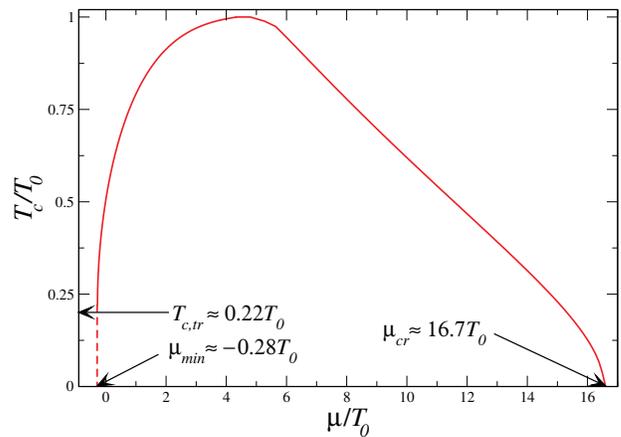}
\end{center}
\caption{FM phase-transition temperature $T_c$ vs. the chemical potential $%
\protect\mu$ (both in units of $T_0$) in a 2D DMS with $S=5/2$. The
abscissa axis also corresponds to a ratio between the free-carrier and magnetic
ion densities $n_e$ and $n_i$ (except a
constant). In the mean-field treatment, $T_c$ would stay at $T_0$ with no
dependence on $\mu$ in the highly degenerate regime. The dashed line
projects the condition at which the FM ordering ceases to exist in the
nondegenerate regime. }
\label{fig:fi2}
\end{figure}

Figure \ref{fig:fi2} illustrates the combined outcome of $T_c$ vs.\
$\mu$ in the full range of $\mu$ (with $S=5/2$).
Evidently, the maximal $T_c$ (or its close vicinity) can be achieved only
in a relatively narrow range of $\mu$ near the crossover point.
This is different from the mean-field model
which predicts $T_c = T_0$ once $\mu$ becomes sufficiently large.
As $\mu $ is lowered, $T_{c}$ shows a rather rapid but continuous decrease to a
value $\approx 0.22T_{0}$ (denoted as $T_{c,tr}$) corresponding to $\mu
_{min}\approx -0.28T_{0}$ and then the
solution ceases to exist abruptly. This threshold behavior originates
from the minimal density of free carriers needed to mediate the indirect
exchange interaction. The latter restriction qualitatively distinguishes
a 2D case from the 3D one, where $T_{c}^{3d}$ can be arbitrarily small
but does not vanish even at an infinitesimal carrier density \cite{SR2001}.

In the heavily degenerate regime, $T_c$ also reveals a gradual decrease
but this time to zero. The $T_{c}=0$ condition from Eq.\ \eqref{tcu3} gives
the critical value $\mu _{cr}\approx 16.7 T_{0}$. This decay to zero is
because  the averaging over the spatial disorder cancels out for $\mu \geq \mu _{cr}$
due to the rapid oscillations of the RKKY range function.
In fact, the above observation reflects the fact that the ratio $\mu /T_0$ represents the
geometric factor proportional to $(\bar{R}/\bar{r})^2$, where
$\bar{R}$ ($\sim n_{i}^{-1/2}$) is a mean distance between magnetic ions
and $\bar{r}$ ($\sim k_{F}^{-1}\sim \mu ^{-1/2}$) approximates the RKKY oscillation period
shown in Fig. \ref{fig:fi1}. Interestingly, its inverse $(\bar{r}/\bar{ R})^{2}$
can be interpreted as the mean number of magnetic ions interacting coherently with
each other by the dominant FM spin-spin coupling.  As increasing $\mu$
reduces $\bar{r}$ and the number of ions interacting coherently, the FM order
becomes unsustainable beyond a certain critical value (i.e., $\mu _{cr}$).
Combined with the analysis in the non-/weakly degenerate regime discussed earlier,
our model predicts that the FM ordering can be achieved
only for $\mu _{min}<\mu <\mu _{cr}$ with the optimum condition around $
5T_{0}$.

Note that the overall behavior of $T_c(\mu)$ for $\mu \gtrsim 5 T_0$
appears similar to that in 3D bulk samples \cite{semst02}. This is to be
expected judging from the comparable characteristics of the RKKY range
functions (except the magnitude) described in Fig.\ \ref{fig:fi1}. The only
difference is the exact form of the normalization factor $T_0$ which shows
disparate functional dependences in the 2D and 3D cases. On the other hand,
this very difference in $T_0$ illustrates a distinct feature of the 2D
system in the non-/weakly degenerate regime. As the curve $T_c(\mu)$ in the
bulk samples has shown qualitative accord with the experiments and Monte
Carlo simulations in the degenerate regime \cite{wme}, it is reasonable to
anticipate a similar level of agreement in the 2D structures under
consideration. Of course, it should be noted that the current
2D model is limited by consideration of Ising impurity spins as described above.

For numerical estimation of $T_{c}$ in a realistic case, a QW of Cd$_{0.9}$Mn%
$_{0.1}$Te is chosen as a specific example. The values of the relevant
parameters found in the literature \cite{di} are the carrier-ion exchange
constant $I=0.88$ eV and the hole effective mass $m^{\ast }=0.8m_{0}$, where $%
m_{0}$ is the free electron mass. In addition, the 2D hole density $n_{e}$ is
estimated to be  $\approx 10^{11}$ cm$^{-2}$ that leads to  $\mu /T_{0}\cong
8$ at $x_{i}=10\%$. Substituting all these values to the expression predicts
$T_{c}\approx 180$ K, which is a high value for a DMS. This analysis further
suggests that $T_{c}$ can be increased by another 30 \% or so (to $\sim 240$
K) if the free hole density is lowered (\textit{not raised} contrary to the
conventional perception) by about 40 \% to the desired $\mu /T_{0}\approx 5$.
Controlling $n_{e}$ (thus, $\mu $) independent of $n_{i}$ is clearly
possible, which is particularly so in the 2D structures. Note that our
estimation of $T_{c}$ is rather rough as the values of the material parameters are
temperature, pressure and other external stimuli dependent. Nevertheless,
the results strongly indicate that the FM ordering can be achieved even
at/above room temperature when the 2D DMS systems are properly optimized.
For instance, recent ab initio calculations predicted the carrier-ion
exchange constant significantly larger than 1 eV in magnetically doped 2D
transition-metal dichalcogenides along with comparable hole effective masses
\cite{pan, jin}. Hence, it is not unreasonable to expect a substantial
enhancement of $T_{c}$ in these structures, where the modulation of free
carrier concentrations over a wide range can be readily achieved \cite{tmd5}.

\section{Summary and Outlook}

Possible magnetic long-range order in the doped planar semiconductor
structures with Ising impurity spins exhibits a large body
of interesting physical effects, making them promising candidates for
spintronic, electronic and even photovoltaic applications \cite%
{fzds2004,bibes, str}. In this work, we demonstrate that the magnetic
impurities, realizing the Friedel oscillations of the constituent 2D free carrier
gas, may generate room temperature FM order in a host structure. The
conditions suitable to reach the maximum possible $T_{c}$ is elucidated,
which can provide a useful guideline for experimental realization. It is noted that
the onset of ferromagnetism considered here is due only to
the RKKY interaction, while there are evidently other mechanisms (like direct
ferro- or antiferromagnetic exchange between the close pairs of impurities)
that can also promote the appearance of magnetic order in the 2D
semiconductor structures \cite{kossut,cibscalb, dietl1}. Further, there is
another important effect which is present in all 2D structures except
graphene. This effect is related to the synergy between the RKKY indirect
exchange coupling and the Rashba spin-orbit interaction, leading to the
interesting phenomena such as the strong anisotropy in the resulting $J(r)$
\cite{bruno}. These and other higher-order effects are outside the scope of
the current study.

\begin{acknowledgments}
This work was supported, in part, by the National Science Center in Poland as a research project
No.~DEC-2017/27/B/ST3/02881 and by the US Army Research Office  (W911NF-16-1-0472).
\end{acknowledgments}

\end{document}